\documentstyle[12pt]{article}

0

\begin{document}


\def\II{\relax{\rm 1\kern-.35em1}}
\def\IP{\relax{\rm I\kern-.18emP}}
\renewcommand{\theequation}{\thesection.\arabic{equation}}
\csname@addtoreset\endcsname{equation}{section}

\begin{flushright}
NSF-ITP-96-75 \\
hep-th/9608104
\end{flushright}
\vglue 1cm

\begin{center}

{\large \bf  $K3$-Fibrations and Softly Broken $N\!=\!4$
Supersymmetric \protect \\

\vspace{5mm}

Gauge Theories.}

\end{center}

\vspace{15 mm}

\begin{center}

C\'{e}sar G\'{o}mez$^a$, Rafael Hern\'{a}ndez$^{a,b}$ and
Esperanza L\'{o}pez$^c$

\vspace{8mm}

$^a${\em Instituto de Matem\'{a}ticas y F\'{\i}sica Fundamental,
CSIC, \protect \\ Serrano 123, 28006 Madrid, Spain}

\vspace{5mm}

$^b${\em Departamento de F\'{\i}sica Te\'{o}rica, C-XI,
Universidad Aut\'{o}noma de Madrid, \protect \\
Cantoblanco, 28049 Madrid, Spain}

\vspace{5mm}

$^c${\em Institute for Theoretical Physics, University of
California, \protect \\ Santa Barbara, CA 93106-4030}

\end{center}

\vspace{3.5cm}

Global geometry of $K3$-fibration Calabi-Yau threefolds, with Hodge
number
$h_{2,1}=r+1$, is used to define $N\!=\!4$ softly broken $SU(r+1)$
gauge
theories, with the bare coupling constant given by the dual heterotic
dilaton, and the mass of the adjoint hypermultiplet given by the
heterotic
string tension. The $U(r+1)$ Donagi-Witten integrable model is also
derived from the $K3$-fibration structure, with the extra $U(1)$
associated to the heterotic dilaton. The case of $SU(2)$ gauge group
is
analyzed in detail. String physics beyond the heterotic point
particle
limit is partially described by the $N\!=\!4$ softly broken theory.

\pagebreak


\section{Introduction}

The Seiberg-Witten solution \cite{SW}-\cite{8} of $N\!=\!2$
supersymmetric gauge theories turns out to be intimately related to
both
integrable models \cite{in}-\cite{9} and string theory
\cite{KV}-\cite{KLMVW}. The mass formula for BPS states
\begin{equation}
M = \sum_{i=1}^{r} \| n_{i}^{e} a_{i}(\vec{u}) +
n_{i}^{m} a_{i}^{D}(\vec{u}) \|,
\end{equation}
where $r$ is the rank of the gauge group of the theory, admits a
geometrical
representation in terms of the periods,
\begin{equation}
a_{i}(\vec{u}) = \oint_{\alpha_i} \lambda, \: \: \: \: \: \: \: \:
a_{i}^{D}(\vec{u}) = \oint_{\beta_i} \lambda,
\label{eq:I2}
\end{equation}
of a meromorphic form $\lambda$ on an hyperelliptic curve of genus r,
$\Sigma_{\vec{u}}$, with $\vec{u}=u_{1}, \ldots ,u_{r}$ and $u_i$ the
Casimir expectation values.

The solution given by (\ref{eq:I2}) defines a family of abelian
varieties,
i.e. the Jacobian of $\Sigma_{\vec{u}}$, parameterized by the
quantum
moduli
manifold. This is the generic structure underlying algebraic
integrable
models \cite{DM}. As it was shown in \cite{DW}, an integrable model
describing $N\!=\!2$ gauge theories can be directly derived, through
Hitchin's construction \cite{H}, from a two dimensional Higgs system
defined on an elliptic Riemann surface, $E_{\tau}$. This approach
leads to
the Seiberg-Witten solution for $N\!=\!2$ gauge theories with one
massive
hypermultiplet in the adjoint representation. The pure gauge theory
appears
as a double scaling limit in which hypermultiplet is decoupled $(m
\rightarrow \infty )$, and the bare coupling constant is sent to
$\infty$.

In addition, the geometrical representation in terms of periods
(\ref{eq:I2}) of the BPS mass formula, calls for a stringy
interpretation
of Seiberg-Witten geometry based on some non critical string, with
effective string tension $\lambda$, winding around an ``internal''
space
described by the Riemann surface $\Sigma_{\vec{u}}$ \cite{KLMVW}.
This
string interpretation nicely combines two ingredients, namely the
string
derivation of Seiberg-Witten solution, and its integrable model
representation.
The first one arises from the discovery of heterotic-type II dual
pairs
\cite{HT,svd,KV,SQ}, with the heterotic string compactified on $K3
\times
T^2$. Higgsing the gauge group leads to different heterotic
compactifications
having $r+2$ vector excitations and $s$ neutral hypermultiplets,
therefore
the corresponding type IIA dual should be defined on a Calabi-Yau
threefold
with $h_{1,1}=r+1$, and $h_{2,1}=s-1$. These Calabi-Yau threefolds
provide
the string theory extension of Seiberg-Witten quantum moduli for a
gauge
group of rank $r$. An important step in the construction of dual
pairs is the
use of threefolds which are $K3$-fibrations \cite{K3}. The reason for
this
requirement goes back to the identification of the Calabi-Yau modulus
that
we could put in correspondence with the heterotic dilaton. This
moduli is
singled out, for $K3$-fibrations, as the one associated with the size
of
the $\IP^{1}$ base space \cite{AL}.

On the other hand, the integrable model representation of the
Seiberg-Witten curve for pure gauge theories,
can be defined as a fibration of the spectral cover set given by the
vanishing locus of the Landau-Ginzburg potential associated to the
Dynkin
diagram of the corresponding gauge group \cite{MW}. The relation
between
this $0$-dimensional fibration, and the $K3$-fibration defining the
Calabi-Yau threefold can be now obtained by fixing the $K3$-fiber at
an
orbifold point, described by the corresponding Dynkin diagram, and
blowing
it up to an ALE space \cite{KLMVW}. This
geometrical manipulation can be formally undertaken by turning
gravity
off $(S \rightarrow \infty)$ and simultaneously going to the point
particle regime by sending the string tension to zero \cite{KKLMV}.
Within this approach, a string interpretation of the periods
(\ref{eq:I2}) is obtained as the wrapping of a $3$-brane on
a $3$-cycle of the Calabi-Yau threefold. To put it more precisely,
the meromorphic form $\lambda$ appears as the string tension of
the self-dual string obtained when wrapping a $3$-brane on $2$-cycles
of
the ALE-space.

In this paper, we will try to go one step further in
the study of the deep interplay between $K3$-fibered Calabi-Yau
threefolds
and the integrability underlying the Seiberg-Witten solution. In
order to
do that we will choose, continuing the line in reference \cite{GHL},
the
approach to integrability used by Donagi and
Witten \cite{DW}, based on Hitchin's gauge model on $E_{\tau}$, where
one
naturally lands onto the Seiberg-Witten solution for
$N\!=\!4$ softly broken gauge theories.
The physical reason for such a choice is related to the existence,
in addition to the Higgs moduli, of two extra parameters, namely,
the bare coupling constant $\tau$
and the bare mass for the hypermultiplet in the adjoint
representation.
Our main goal will be to find a string derivation of this theory.
Moreover
we will claim that $K3$-fibered threefolds naturally
lead to the $N\!=\!4$ softly broken version of $N\!=\!2$ gauge
theories,
with the
bare coupling constant $\tau$ and the hypermultiplet mass mapped in a
concrete
way into the heterotic dilaton and the string scale $\alpha'$
respectively.

The main difference between our case and the string derivation of
pure
$N\!=\!2$ gauge theory is
that, in order to obtain the $N\!=\!4$ softly broken version, we
should
work beyond the point particle limit, at generic values of the
heterotic
dilaton field. Thus, we will be using the global information on the
$K3$-fibration structure. The important physical question raised by
this
analysis
is, of course, to unravel the type of global string dynamics that we
are
capturing with the $N\!=\!4$ softly broken quantum field theory.
We will concentrate mainly in the case of gauge group $SU(2)$.
What we observe is that the gauge theory
that we are associating with the threefold fibration captures the
features
which are universal for a certain set of $K3$-fibrations, with
identical
Hodge number $h_{2,1}=2$, but differing in the modular properties of
the
mirror map and in Hodge number $h_{1,1}$.
In particular, following \cite{AP}, we will relate the different 
modular properties of the mirror map to the Kac-Moody level at which 
the gauge symmetry is realized in the heterotic dual string.
The $N=4$ softly broken theory will then determine the
relation between the Kac-Moody level and the genus of a curve of 
singularities developed by the Calabi-Yau manifolds at the locus
corresponding to vanishing heterotic dilaton \cite{KM,KMP}.

The $N\!=\!4$ softly broken version of the Calabi-Yau threefold
singles out a particular singular locus on the moduli of complex
structures of the threefold, namely that in correspondence with the
field
theory locus where some component of the adjoint hypermultiplet
becomes
massless. If the mass of the hypermultiplet is correctly capturing
part of
the dynamics of the string scale, $\alpha'$, we should expect the
corresponding singular locus of the Calabi-Yau threefold to be
somehow
related to the existence of the string scale, in very much a similar
way as the self dual point for compactifications on $S^{1}$ is given
by a
radius $R$ equal to $\sqrt{\alpha'}$. We will present some evidence
in this
direction.

Concerning Donagi-Witten version of integrability as based on
Hitchin's
gauge model on an elliptic curve $E_{\tau}$, we will observe that
this
``reference Riemann surface'' can also be recovered from the
geometrical
data of the $K3$-fibration.


\section{The Calabi-Yau Curve and its Quantum Field Theory Analogue.}

\subsection{$K3$-Fibrations.}

Let us consider the string embedding of $SU(2)$ $N=2$
supersymmetric Yang-Mills theory according to \cite{KV}. The Higgs
mechanism that produces the desired gauge group originates 129
neutral
hypermultiplets, therefore we must
choose as the type IIA dual a $K3$-fibered threefold with Hodge
numbers $h_{1,1}=2$, $h_{2,1}=128$, i.e.
$W=\IP_{\{1,1,2,2,6\}}^{4}[12]$.
The mirror manifold $W^*$ can be obtained from the orbifold
construction \cite{GP}, and has defining polynomial
\begin{equation}
W^{\ast} = \frac{1}{12}  x_{1}^{12} + \frac{1}{12}  x_{2}^{12} +
\frac{1}{6} x_{3}^6 + \frac{1}{6} x_{4}^6 + \frac{1}{2} x_{5}^2
- \psi  x_1 x_2 x_3 x_4 x_5 - \frac{1}{6} \phi (x_1 x_2)^6.
\label{1000}
\end{equation}
The moduli space of complex deformations of $W^*$ is parameterized by
$(\psi,\phi)$, subject to the global symmetry
\begin{equation}
{\cal A}:(\psi,\phi) \rightarrow (\beta \psi, - \phi)~, \hspace{1cm}
\beta^{12}=1.
\label{555}
\end{equation}
This symmetry forces to introduce invariant quantities; we will use
$b=1/\phi^2$ and $c=-\phi / \psi^6$.
The $K3$-fibration structure of (\ref{1000}) becomes manifest by the
change
of variables $x_1/x_2 \equiv z^{1/6}b^{-1/12}$,
$x_{1}^2 \equiv x_0 z^{1/6}$ \cite{KLMVW}:
\begin{equation}
W^{\ast}= \frac{1}{12}(z+\frac{b}{z}+2) x_{0}^6 + \frac{1}{6} x_{3}^6
+
\frac{1}{6} x_{4}^6 + \frac{1}{2} x_{5}^2 + c^{-1/6} x_0  x_3 x_4
x_5,
\label{2}
\end{equation}
with the variable $z$ acting as coordinate on the $\IP^1$ base space.
It is convenient to define
\begin{equation}
d(z;b)=\frac{1}{2}(z+\frac{b}{z}+2)~, \hspace{1cm}
\hat{c}(z;b,c)=c \: d(z;b).
\label{557}
\end{equation}
Substituting this into (\ref{2}) and rescaling $x_0$, $W^{\ast}$
acquires
the explicit form of a $K3$-surface
\begin{equation}
W^{\ast} = \frac{1}{6} x_{0}^6 + \frac{1}{6} x_{3}^6 +
\frac{1}{6} x_{4}^6 + \frac{1}{2} x_{5}^2 +
\hat{c}(z;b,c)^{-1/6} x_0  x_3 x_4 x_5.
\label{1001}
\end{equation}
As we move in $\IP^1$, the $K3$-fiber can become singular. From
(\ref{1001}) it is easy to deduce that this occurs for the $K3$
modulus values $\hat{c}(z;b,c)=0,1$. These values of $\hat{c}$
are acquired at the following $\IP^1$ points, $z=e_{i}^{\pm}$:
\begin{eqnarray}
\label{3}
\hat{c}=0 & \rightarrow & e^{\pm}_0 =  -1 \pm \sqrt{1-b},  \nonumber
\\
\hat{c}=1 & \rightarrow & e^{\pm}_1 = \frac{1 - c \pm
\sqrt{(1-c)^2 - b c^2}}{c}.
\end{eqnarray}
The discriminant of (\ref{1001}) is therefore given by
$\Delta(z;b,c)=
\prod_{i=0}^{1} (z-e^{+}_i(b,c))(z-e^{-}_i(b,c))$. There is an
additional
singularity at $\hat{c}(z;b,c)=\infty$, which is originated in the
quotient by discrete reparameterizations of (\ref{1001}) inherited
from
the orbifold construction of $W^{\ast}$. It corresponds to the points
\begin{eqnarray}
\hat{c}=\infty & \rightarrow & z=0,\infty \hspace{1cm}
(b \neq \infty).
\label{560}
\end{eqnarray}

The Calabi-Yau manifold becomes singular when some of the points
(\ref{3})-(\ref{560}) coalesce. We will now analyze the regions in
moduli
space where this situation happens \cite{can,CYY}
(we will follow notation in \cite{can}).
The loci
\begin{eqnarray}
\label{l1}
       {\cal C}_1 & = & \{ b=1 \}, \nonumber \\
       {\cal C}_C & = & \{(1-c)^2-bc^2=0\} ,
\label{eq:c4}
\end{eqnarray}
are respectively obtained from the identifications $e_0^+=e_0^-$ and
$e_1^+=e_1^-$. ${\cal C}_C$ is the conifold locus, where
$3$-cycles of the threefold degenerate to points, while  ${\cal C}_1$
corresponds to the appearance of a genus two curve of $A_1$
singularities.
We can also consider
\begin{eqnarray}
       {\cal C}_0 & = & \{ c = \infty \}, \nonumber \\
       {\cal C}_{\infty} & = & \{ b = 0 \},
\label{eq:c5}
\end{eqnarray}
which are defined, respectively, by the identifications
$e_1^{\pm}=e_0^{\pm}$
and $e_0^+=e_1^+=0$. ${\cal C}_0$ is an orbifold locus, given
by the fixed points under ${\cal A}^2$. ${\cal C}_{\infty}$
corresponds to
the weak coupling limit locus, once we identify the heterotic dilaton
with
the size of the base space \cite{AL}. In addition we have
\begin{equation}
       {\cal D}_{(0,-1)}  =  \{ c = 0 \},
\label{l2}
\end{equation}
implying $e_{1}^+=0$, $e_{1}^-=\infty$. ${\cal D}_{(0,-1)}$ is an
exceptional divisor in moduli space whose intersection with
${\cal C}_{\infty}$ identifies the large complex structures limit of
(\ref{1000}).
Finally, let us notice that at $b=\infty$ the points (\ref{560}) are
ill defined, giving raise to a very degenerate situation. We can put
in correspondence
\begin{equation}
       {\cal D}_{(-1,0)}  =  \{ b = \infty \},
\label{556}
\end{equation}
with the exceptional divisor introduced to resolve a conical
singularity
in the moduli space generated from quotienting by the discrete
transformation ${\cal A}$\footnote{Defining the ${\cal A}$-invariant
quantities $\xi \equiv \psi^8$, $\eta \equiv \psi^4 \phi$ and $\zeta
\equiv \phi^2$, the quotiented moduli space is given by the
projective cone $\xi \zeta = \eta^2$.}.
In section 2.3 we will return again
to this point, in relation with a double covering of the moduli
space\footnote{This will explain why the analysis of singular loci
provided by the fibration structure is missing the
${\cal D}_{(-1,-1)}$ divisor, which, together with ${\cal
D}_{(0,-1)}$,
is associated with the resolution of a tangency point between
${\cal C}_{\infty}$ and ${\cal C}_1$ originated in the ${\cal A}$
quotient. This completes the set of toric divisors in the
compactification
of the moduli space worked out in \cite{can}.}.

More in general, we can consider $K3$-fibered Calabi-Yau threefolds
whose
mirror $W^*$ is also a $K3$-fibration that can be written as
\cite{KLMVW}
\begin{equation}
W^{\ast}=\frac{1}{2n}\left( z + \frac{b}{z} +2 \right) x_{0}^n +
W_{K3}^{\ast} (x_0,x_3,x_4,x_5;c_i).
\label{999}
\end{equation}
The discriminant of the fiber will be given by a polynomial
$\Delta(z;b,c_i)$, depending on the point $z$ in the base and the
moduli
parameters, with zeroes at $z=e_i^{\pm}$ where the $K3$-fiber becomes
singular. The number of singular points is given by
$2 h_{2,1}=2 (r+1)$.

\subsection{The Calabi-Yau Curve.}

{}From the above $K3$-fibration structure it is possible to 
recover \cite{KLMVW}, in the
heterotic point particle limit $b \rightarrow 0$, $\alpha'
\rightarrow 0$
\cite{KKLMV}, the Seiberg-Witten curves for $N\!=\!2$ $SU(r+1)$
Yang-Mills theory \cite{KLTY,AF,MW}. 
Namely for the $SU(2)$ case, using the map
\begin{eqnarray}
b & = & \alpha'^{2} \Lambda^{4}, \nonumber\\
c & = & 1 + \alpha' u,
\label{eq:cc1}
\end{eqnarray}
with $\Lambda$ the $SU(2)$ dynamical scale, and rescaling
$z \rightarrow \alpha' z$, the points (\ref{3})
become the branch points of the associated Seiberg-Witten
curve \cite{MW}
\begin{eqnarray}
e_0^{\pm} & = & 0, \infty , \nonumber\\
e_1^{\pm} & = & - u \pm \sqrt{u^2-\Lambda^4}.
\label{888}
\end{eqnarray}
In general, the geometrical meaning of the point particle limit
amounts
to replacing the $K3$-fiber by an ALE space that blows up an $A_r$
orbifold
singular $K3$. Let us denote by $G=SU(r+1)$ the gauge group whose
Dynkin
diagram describes the orbifold singularity. In this situation, the
branched
cover of the
$\IP^1$ base space that defines the Seiberg-Witten curve
for the pure gauge theory in the integrable model formulation of
\cite{MW} (whose branch points are at the generalization of
(\ref{888})), ${\cal C}_G$,  can be directly derived from the ALE
space homology.

A different approach consists in using the global structure of the
$K3$-fibration in order to build an hyperelliptic curve such that its
moduli space is isomorphic to the Calabi-Yau moduli. We can define
\begin{equation}
{\cal C}^{CY}: \hspace{5mm} y^2=\Delta(z;b,c_i)=\prod_{i=0}^{r}
(z-e^{+}_i(b,c_i))(z-e^{-}_i(b,c_i)).
\label{eq:c13}
\end{equation}
By construction, this curve becomes singular at the moduli values
where
the threefold $W^*$ acquires a singularity, i.e. those where two
roots coalesce. Performing
the point particle limit \cite{KKLMV}, ${\cal C}^{CY}$ reduces
to the Seiberg-Witten curve ${\cal C}_G$.
However, it is important to stress the following.
In order to define ${\cal C}^{CY}$ we have interpreted the singular
points
of the $K3$-fibration as branch points of an associated hyperelliptic
curve.
But contrary to the point particle limit, in which the $K3$-fiber is
substituted by an ALE space, we are not using the $K3$ periods for
the direct
construction of the curve.

Our first aim in analyzing (\ref{eq:c13}) will be, following
previous work in reference \cite{GHL}, to identify the
$N=2$ supersymmetric gauge theory represented by ${\cal C}^{CY}$.
Being defined for arbitrary values of the moduli parameters, the
first difference between ${\cal C}_G$ and ${\cal C}^{CY}$
is that the second depends on the ``dilaton modulus'' $b$.
In the heterotic string framework, the expectation value of the
dilaton field determines the bare gauge coupling constant.
Therefore we should look for a gauge theory in which the coupling
constant behaves as a modulus, i.e. an ultraviolet finite theory.

The simplest candidate for ${\cal C}^{CY}$ is a theory with the field
content of $N=4$, namely Yang-Mills plus a matter hypermultiplet in
the adjoint representation of $G$, where in general the adjoint
hypermultiplet can be massive. Let us consider the case where the
adjoint hypermultiplet has a bare mass $m$. It was shown in
\cite{SW2}
that in a double scaling limit that sends $m$ and the $N=4$ coupling
constant to $\infty$, we can recover the pure Yang-Mills theory.
In this process the mass plays a role formally analogous to that of
the string coupling constant in the double limit that takes
${\cal C}^{CY}$ to ${\cal C}_G$. Therefore, in the proposed
interpretation of string notions in terms of gauge theory quantities,
we should identify $(\alpha')^{-1} \sim m^2$.

Let us now review briefly some results in reference \cite{GHL} for
the case $G=SU(2)$. The curve for $SU(2)$ $N=2$ gauge theory with
one massive adjoint hypermultiplet \cite{SW2}, is given by
\begin{equation}
y^2= (x - a_1 \hat{u} + a_{2}^2)(x  + a_2 (\hat{u} - a_1))
(x - a_2 (\hat{u} - a_1)),
\label{11}
\end{equation}
with $u= \mbox {Tr} \phi^2$ the quadratic Casimir, and $\hat u$
defined by the convenient normalization $\frac{1}{4}m^2 {\hat u}= u$.
The quantities $a_1$ and $a_2$ depend on the asymptotic value of
the gauge coupling constant of the theory, $\tau=\frac{\theta}{2\pi}+
i \frac{4 \pi}{g^2}$, according to\footnote{The Weierstrass
invariants
$e_{i}$ can be defined in the terms of Jacobi theta functions:
$e_{1}=\frac
{1}{3}(\theta_{2}^{4}(0,\tau)+\theta_{3}^{4}(0,\tau)),~~
e_{2}=- \frac
{1}{3}(\theta_{1}^{4}(0,\tau)+\theta_{3}^{4}(0,\tau)),~~
e_{3}=\frac {1}{3}(\theta_{1}^{4}(0,\tau)-\theta_{2}^{4}(0,\tau))$.}
\begin{equation}
a_1= \frac{3}{2}e_1(\tau)~~, \hspace{1cm} a_2= \frac{1}{2} (e_3(\tau)
- e_2(\tau)).
\end{equation}
In the moduli space $(\hat{u}, \tau)$, we can differentiate the
following loci where (\ref{11}) becomes singular
\begin{equation}
	\begin{array}{lcl}
	\hat{{\cal C}}_{0}  & = & \{ \hat{u}(\tau)= a_1(\tau) \}, \\
	\hat{{\cal C}}_{C}^{(1)} & =
	& \{ \hat{u}(\tau)= a_2 (\tau) \}, \\
	\hat{{\cal C}}_{C}^{(2)} & = & \{ \hat{u}(\tau)= -a_2(\tau) \}, \\
	\hat{{\cal D}} & = & \{ \hat{u} = \infty \}.
	\end{array}
\hspace{1cm}
	\begin{array}{lcl}
	\hat{{\cal C}}_{\infty}  & = & \{ \tau= i \infty \}, \\
	\hat{{\cal C}}_{1}^+     & = & \{\tau = 0 \}, \\
	\hat{{\cal C}}_{1}^-     & = & \{\tau = 1 \}, \\
	&&
	\end{array}
\label{12}
\end{equation}

We can now try to put in correspondence the moduli space of K\"ahler
deformations of $\IP^{4}_{\{1,1,2,2,6\}}[12]$ with the moduli space
of the
$N=4$ softly broken theory (\ref{11}). The basic idea will be to
map the singular loci described by the fibration structure of the
mirror
$W^{\ast}$, (\ref{l1})-(\ref{l2}), with the set (\ref{12}).
This is achieved by the map \cite{GHL}
\begin{equation}
c=\frac{a_1(\tau)}{a_1(\tau)-\hat{u}}~~, \hspace{1cm}
b=\left( \frac{a_2 (\tau)}{a_1(\tau)} \right)^2.
\label{14}
\end{equation}
It is important to notice, using the modular properties of
$a_i(\tau)$\footnote{We have $a_1(\tau+1)=a_1(\tau)$ and
$a_2(\tau+1)=
-a_2(\tau)$.},  that (\ref{14}) is effectively quotienting
by the $(\hat{u},\tau)$-plane transformation
\begin{equation}
T:(\hat{u},\tau) \rightarrow (\hat{u},\tau +1).
\label{eq:c11}
\end{equation}
Indeed the proposed map sends $\hat{{\cal C}}_{C}^{(1,2)} \rightarrow
{\cal C}_{C}$ and $\hat{{\cal C}}^{\pm}_1 \rightarrow {\cal C}_1$,
while $\hat{{\cal C}}_{\infty}$, $\hat{{\cal C}}_{0}$ and $\hat{{\cal
D}}$,
which are fixed under (\ref{eq:c11}), are mapped respectively into
${\cal C}_{\infty}$, ${\cal C}_0$ and ${\cal D}_{(0,-1)}$\footnote{A
similar map
between the singular loci of the hyperelliptic curve describing
$SU(3)$
$N\!=\!2$ supersymmetric Yang-Mills theory with adjoint matter and
the moduli 
space of the $SU(3)$ gauge group Calabi-Yau manifold \cite{KV},
$\IP_{\{1,1,2,8,12\}}^{4}[24]$, is proposed in Appendix B.}.
Therefore we observe that the $(\hat{u},\tau)$-plane behaves
as
a double cover of the Calabi-Yau moduli space.

In the weak coupling limit $\tau \rightarrow i \infty$, the map
(\ref{14}) becomes
\begin{equation}
c=\frac{1}{1-\hat{u}}~~, \hspace{1cm}
b=64 e^{2 \pi i\tau}.
\label{15}
\end{equation}
{}From the heterotic-type II identification at leading order
$b=e^{-S}$
\cite{KV}, with $S$ the heterotic dilaton, the above expression
explicitly
shows that we are associating $\tau$ with $S$.
Setting $(\alpha')^{-1}=\frac{1}{4}m^2$, as we proposed, (\ref{15})
reproduces the relation $\frac{c^2 b}{(1-c)^2}=\frac{\Lambda^4}{u^2}$
used in \cite{KKLMV} for defining the point particle limit of the
string
(of which (\ref{eq:cc1}) is the first order).

We would like now to prove that, by the map (\ref{14}), the curve
${\cal C}^{CY}$ is in fact the Seiberg-Witten curve for
$N\!=\!2$ $SU(2)$ Yang-Mills theory with one massive hypermultiplet
in the adjoint representation. In order to do so, we rewrite the
Calabi-Yau curve (\ref{eq:c13}) for the $SU(2)$ case as
\begin{equation}
y^2= \prod_{i=0,1} (z-e_{i}^+(\hat{u},\tau))(z-e_{i}^-
(\hat{u},\tau)),
\label{16}
\end{equation}
with
\begin{eqnarray}
\label{510}
e_{0}^{\pm} & = & \frac{-a_1(\tau) \pm \sqrt{ a_1(\tau)^2 -
a_2(\tau)^2}}{ a_1(\tau)}, \nonumber \\
e_{1}^{\pm} & = & \frac{- \hat{u} \pm \sqrt{ {\hat u}^2 -
a_2(\tau)^2}}{ a_1(\tau)}.
\end{eqnarray}

It is convenient now to transform the Calabi-Yau curve (\ref{16})
into
the standard cubic form $y^2=z(z-1)(z-\lambda)$, where
\begin{equation}
\lambda=\frac{(e_{1}^{-} - e_{0}^+)(e_{0}^{-} -
e_{1}^+)}{(e_{1}^{+} - e_{0}^+)(e_{0}^{-} - e_{1}^-)}.
\label{17}
\end{equation}
Using results in Appendix A, we observe that the field theory curve
(\ref{11}) is isogenic to the quartic
\begin{equation}
y^2=(x^2 + a_1 \hat{u} - a_{2}^2)^2 - (a_2(\hat{u}-a_1))^2.
\label{20}
\end{equation}
The $\lambda$ parameter that characterizes the standard cubic form
of (\ref{20}) is given by
\begin{equation}
\lambda'= \frac{a_1 \hat{u}-a_{2}^2 + \sqrt{(a_{1}^2 -
a_{2}^2)(\hat{u}^2 - a_{2}^2)}}{a_1 \hat{u} - a_{2}^2 -
\sqrt{(a_{1}^2 -
a_{2}^2)(\hat{u}^2 - a_{2}^2)}}.
\label{511}
\end{equation}
Substituting (\ref{510}), we can see that $\lambda$ and $\lambda'$
precisely coincide.

This concludes the proof that ${\cal C}^{CY}$ for the Calabi-Yau
mirror
of $\IP_{\{1,1,2,2,6\}}^{4}[12]$ is, by the map (\ref{14}), the
curve describing $SU(2)$ $N=2$ supersymmetric Yang-Mills with one
massive
hypermultiplet in the adjoint representation. The necessity of
introducing
an isogeny transformation is originated in different conventions for
the
Higgs field normalization. The curve (\ref{11}) follows the
convention
in \cite{SW}, which is appropriate when only integer charges for BPS
states can appear.
However the Seiberg-Witten curve ${\cal C}_G$, obtained from the
point
particle limit of the string, adopts a normalization
adequated to gauge theories that can include fundamental matter
\cite{SW2}.
Since ${\cal C}^{CY}$ flows to ${\cal C}_G$ in the point particle
limit,
it shares with it the same normalization of the Higgs field,
differing,
up to an isogenic transformation (see (\ref{A1})), of that in
(\ref{11}).

\subsection{Double Covering and $K3$-Fibrations.}

We have pointed out that the $(\hat{u},\tau)$-moduli space of the
$SU(2)$ theory with adjoint matter, using the map (\ref{14}),
acts as a double covering of the Calabi-Yau $\IP_{\{1,1,2,2,6\}}[12]$
moduli space $(b,c)$. At the same time, we have seen that
the Seiberg-Witten curve for this gauge theory can be explicitly
constructed from string theory compactification. This
raises the following puzzle: if the string moduli space
is correctly labeled by $(b,c)$, how can the double covering
variables $(\hat{u},\tau)$ be naturally derived from
the Calabi-Yau moduli?

The answer comes from the way we define the $K_3$-fibration.
In (\ref{2}) the choice $\sqrt{b}=- \frac {1}{\phi}$ has been
implicitly
done. If we instead choose $\sqrt{b}= \frac {1}{\phi}$, the fibration
structure (\ref{557}) changes into
\begin{equation}
\hat{c}~'= \frac{1}{2}\left(-z-\frac{b}{z} +2 \right)c.
\label{eq:c20}
\end{equation}
We see that the choice of one or another branch of
$\sqrt{b}$ amounts to changing $z \rightarrow -z$. This is an effect
that can be absorbed in a trivial redefinition of $z$, and therefore
does not,
in essence, affect ${\cal C}^{CY}$. However it shows how the
$K_3$-fibration
structure, and any notion based on it, leads to a double covering of
the
moduli space determined by the two branches of $\sqrt{b}$. Moreover
in
terms of the map (\ref{14}), written as
\begin{equation}
\sqrt{b}=\frac{a_2(\tau)}{a_1(\tau)},
\label{200}
\end{equation}
the change of branch becomes equivalent to the transformation
$T: \tau \rightarrow \tau +1$.

This problem is related to a subtlety in the derivation of the
Seiberg-Witten solution for $SU(2)$ Yang-Mills in reference
\cite{KKLMV},
accomplished by performing
the blow up of the tangency point $(b=0,c=1)$ between the conifold
locus
${\cal C}_C$ and the weak coupling locus ${\cal C}_{\infty}$.
The variable parameterizing the second exceptional divisor arising
from the
blow up of the tangency, $\frac {bc^2}{(1-c)^2}$, is identified with
the field theory {\bf Z}$_2$ invariant quantity $\frac
{\Lambda^4}{u^2}$.
The reason why the blow up approach recovers $(\Lambda^2/u)^2$
is that, from (\ref{eq:cc1}), the change of branch
$\sqrt{b} \rightarrow -\sqrt{b}$ implies $\Lambda^2 \rightarrow
-\Lambda^2$,
and the Calabi-Yau moduli space naturally quotients by this
transformation.
On the contrary, from the $K3$-fibration
in the ALE limit we get in a direct way the Seiberg-Witten curve,
parameterized by $u$, and not its {\bf Z}$_2$ quotient.

The transformation $T$ has indeed an string analogue. Using the
map (\ref{14}), $T$ is equivalent to the discrete symmetry
(\ref{555})
of the Calabi-Yau moduli space, ${\cal A}$. After performing the
${\cal A}$-quotient, the Calabi-Yau moduli space becomes isomorphic
\cite{can} to the space $\IP^{2}_{\{1,1,2\}}$, which has a conical
singularity at the origin (see footnote 1). This is precisely the
geometry
of the $T$-quotiented $(\hat{u},\tau)$-plane\footnote{Defining the
invariants with respect to (\ref{eq:c11}) $\xi \equiv \tilde{u}^2$,
$\eta \equiv \frac {\tilde{u}}{\epsilon}$, $\zeta \equiv
\frac {1}{\epsilon^2}$, where $\epsilon \equiv 8 e^{\pi i \tau}$,
$\tilde{u} \equiv \frac {\hat{u}} {\epsilon}$, the quotiented
$(\hat{u},\tau)$-moduli space is also given by the projective cone
$\xi \zeta = \eta^2$.}. We stress that the
$(\hat{u},\tau)$ variables, although
undo the ${\cal A}$-quotient, preserve an ${\cal A}^2$-quotient.
This can be seen from the fact that ${\cal A}^2$ fixes the locus
${\cal C}_0$, which has a counterpart in the $(\hat{u},\tau)$ moduli
space.

Finally we notice that $(\hat{u},\tau)$ points are
not in a one-to-one correspondence with $(\sqrt{b},c)$. The divisor
${\cal D}_{(-1,0)}=\{ b=\infty \}$, which appears in the
blow up of the conical singularity created by the ${\cal
A}$-quotient,
presents non-trivial monodromy. However in the
$(\hat{u},\tau)$-plane, having undone the quotient by ${\cal A}$, the
value $b=\infty$ should not imply an additional monodromy locus.
This is in fact the case. Since $b=\infty$ corresponds to
$a_1=0$, from (\ref{14}) we see that
${\cal D}_{(-1,0)}$ is blown down to the point $\hat{u}=0$.
The fibration structure is also reflecting this remark through its
dependence on the $K3$-fiber modulus $\hat{c}=cd$.
The value $b=\infty$ implies $d=\infty$, so that the combination
$\hat{c}$ blows down the line $(b=\infty,c)$ to a point.

\section{Integrability.}

In the previous section we have compared ${\cal C}^{CY}$ with the
curve
for $SU(2)$ $N=2$ Yang-Mills with one massive hypermultiplet in the
adjoint
representation. In this section we will recover, from the Calabi-Yau
manifold, the basic building elements used in the Donagi-Witten
formulation
\cite{DW} of the integrability of gauge theories with adjoint matter.

To start with, let us just briefly recall the construction in
\cite{DW} for
the simple case of $SU(2)$ gauge theory. According to Hitchin's
construction,
we start with the elliptic curve
\begin{equation}
E_{\tau}: \: \:  y^2 = (x-e_1(\tau))(x-e_2(\tau))(x-e_3(\tau)),
\label{eq:i1}
\end{equation}
on which a two dimensional Higgs field $\phi$ is defined as an
holomorphic
1-form transforming in the adjoint representation of the $SU(2)$
gauge group.
In terms of $\phi$, an spectral cover curve is defined through
(\ref{eq:i1})
and
\begin{equation}
0=t^2 -x + A_2,
\label{eq:i2}
\end{equation}
with $A_2$ related to the Higgs expectation value of the $N\!=\!2$
theory
by $A_2 = u - \frac {1}{2}e_1(\tau)$.
Equations (\ref{eq:i1}) and (\ref{eq:i2}) define a genus $2$ curve,
symmetric
under the {\bf Z}$_2 \times${\bf Z}$_2$ transformation $t \rightarrow
-t$,
$y \rightarrow -y$. The curve (\ref{11}) we have employed in the
previous
section corresponds to the {\bf Z}$_2 \times${\bf Z}$_2$ invariant
part.

The integrable model version of pure gauge theory curves
\cite{in,MW}, which is
recovered from string theory in the point particle limit
\cite{KLMVW}, is
given by
\begin{equation}
z + \frac {\Lambda^4}{z} + 2 P_{A_1}(t,u)=0.
\label{eq:curve}
\end{equation}
This can be derived from the ``classical'' expression
(\ref{eq:i2})\footnote{The Landau-Ginzburg potential
$P_{A_1}(t,u)=t^2+u$
is being used.} by the ``quantization'' procedure
$x \rightarrow -\frac{1}{2}(z + \frac {\Lambda^4}{z})$. Instead, in
\cite{DW}
the quantization of (\ref{eq:i2}) is
implemented by forcing $x$ to live on the elliptic curve
(\ref{eq:i1}),
parameterized by the $N=4$ coupling constant $\tau$.

In order to show that the Donagi-Witten integrability construction is
the
natural extension of (\ref{eq:curve}) when we move off the point
particle
limit, we have still to determine the elliptic curve $E_{\tau}$
from Calabi-Yau data. Associated to a $K3$-fibration threefold
\begin{equation}
W^{\ast}=\frac{1}{n} d(z;b)~ x_{0}^n + W_{K3}^{\ast}
(x_0,x_3,x_4,x_5;c_i)
\label{100}
\end{equation}
with $d(z;b)$ given by (\ref{557}), we can always consider the
following
four points on the $\IP^1$ base space where the $K3$-fiber becomes
singular
\begin{equation}
\begin{array}{ccc}
d(z;b)=0 & \rightarrow & e_{0}^{\pm} = -1 \pm \sqrt{1-b}, \\
d(z;b)=\infty & \rightarrow & z=0,~\infty.
\label{eq:i6}
\end{array}
\end{equation}
In the spirit of the previous section, we can define the elliptic
curve
associated to these data
\begin{equation}
y^2=z(z-e_{0}^+)(z-e_{0}^-).
\label{105}
\end{equation}
This curve is independent of the particular Calabi-Yau we are working
with, as far as the $K3$-fibration structure is defined by
$d=\frac{1}{2}(z+\frac{b}{z}+2)$. The curve (\ref{105}) is
characterized by the $\lambda$ parameter
\begin{equation}
\lambda=\frac{e_{0}^-}{e_{0}^+},
\label{110}
\end{equation}
which only depends on the dilaton modulus $b$.

We will follow the same path as in section 2.2 for comparing
(\ref{105})
with $E_{\tau}$. Namely, after shifting $x \rightarrow
x+\frac{1}{2}e_1$, we apply the isogeny described in Appendix A.
This transforms $E_{\tau}$ into the quartic
\begin{equation}
y^2=(x^2 +a_1(\tau))^2 - a_{2}(\tau)^2.
\end{equation}
Using the map between $b$ and $\tau$ proposed in (\ref{14}), the
$\lambda$
parameter of this curve can be easily seen to coincide with
(\ref{110}),
implying that (\ref{105}) is isogenic to $E_{\tau}$. In this sense we
notice that, according to (\ref{A1}), the elliptic parameter of the
curve
(\ref{105}) is $2\tau$.

Let us stress the meaning of the extra singularity at
$d = \infty$ used in (\ref{eq:i6}), which corresponds to
$z=0,\infty$.
The Seiberg-Witten differential for ${\cal C}^{CY}$,
which we will describe in the begining of the next section (see
(\ref{41})),
has poles with residue at the points $z=0,\infty$. The residue at
these
poles is defining the mass of the hypermultiplet in the adjoint
\cite{SW2}.
(In the point particle limit \cite{KKLMV} the structure
(\ref{105}) disappears, as $e_0^{\pm} \rightarrow 0, \infty$).
In the context of the Donagi-Witten formulation, the singularities at
$z=0,\infty$ have a similar role to the point $x=\infty$ in
$E_{\tau}$
which, by equation (\ref{eq:i2}), corresponds to a degenerate
spectral set
$t = \pm \infty$. Recalling the underlying Hitchin model, 
we observe that $x=\infty$ is the pole of the associated two
dimensional Higgs field, whose residue also defines the mass of the
adjoint
hypermultiplet \cite{DW}.

In summary, from the Calabi-Yau geometry we get the integrability
structure
of the Seiberg-Witten model, as it is described in the Donagi-Witten
construction. This integrability structure only shows up when we keep
alive
both the string scale, $\alpha'$, and the gravitational effects due
to the
existence of the dilaton.
However, we can not expect the Donagi-Witten model to be equivalent
to full
fledged string theory, an issue that will be addressed in the next
section.

\section{Picard-Fuchs Equations.}

It was shown in reference \cite{KLMVW} that in the point particle
limit,
where the $K3$ degenerates to an ALE space, we can effectively
map the second homology group of $K3$ into the homology group of
0-cycles defined by the spectral set $\{ t~~ |~~ P_{A_1}(t;u_i)=0
\}$.
Alternatively, the integration of the holomorphic top form of the
Calabi-Yau manifold on a 2-cycle of the $K3$, in its ALE limit,
defines
the meromorphic Seiberg-Witten form $\lambda$ for the curve
(\ref{eq:curve}).
This form is \cite{MW}
\begin{equation}
\lambda= \sqrt{u + \frac{1}{2} (z + \frac{\Lambda^4}{z} )}~ 
\frac{dz}{z}.
\label{40}
\end{equation}
Using the Calabi-Yau curve defined in the previous sections we can
propose a generalization of (\ref{40}) to the case gravity is turned
on. Namely, we can consider the meromorphic form $\lambda$ derived
from the ${\cal C}^{CY}$ by means of the map (\ref{14}). The result
is
\begin{equation}
\tilde{\lambda} = \sqrt{1-\frac{1}{\hat{c}}}~ \frac{dz}{z},
\label{41}
\end{equation}
with $\hat c$ defined in (\ref{557}).
Our aim in this section is to understand the meaning of (\ref{41}),
and equivalently ${\cal C}^{CY}$, in the string context.
In order to do so, we will analyze in what way
$\tilde{\lambda}$ is related with the periods of $K3$ and with
the associated Picard-Fuchs equation.

The information used to construct ${\cal C}^{CY}$ reduces to
the discriminant of the $K3$-fiber $\Delta(z;b,c)$ and,
derived from it, the discriminant of the Calabi-Yau threefold.
However these data do not determine in an unique way the threefold.
There exist different Calabi-Yau spaces whose moduli of K\"ahler
deformations share common features.
As an example the manifold $\IP_{\{1,1,2,2,6\}}^{4}[12]$ we have been
considering up to now, exhibits the same singular loci and Yukawa
couplings structure that the hypersurface
$\IP_{\{1,1,2,2,2\}}^{4}[8]$
and the complete intersections $\IP_{\{1,1,2,2,2,2\}}^{5}[4,6]$ and
$\IP_{\{1,1,2,2,2,2,2\}}^{6}[4,4,4]$ \cite{can,CYY,Y2}.
These four manifolds are $K3$-fibrations and from any of them, in
the point particle limit (\ref{eq:cc1}), can be derived the exact
physics of $SU(2)$ $N=2$ Yang-Mills theory. For simplicity
we will denote them as $A:\IP_{\{1,1,2,2,6\}}^{4}[12]$,
$B:\IP_{\{1,1,2,2,2\}}^{4}[8]$, $C:\IP_{\{1,1,2,2,2,2\}}^{5}[4,6]$
and $D: \IP_{\{1,1,2,2,2,2,2\}}^{6}[4,4,4]$ (see Appendix C for a
brief
description of these spaces).

The Picard-Fuchs equation for their $K3$-fiber is \cite{LY}
\begin{equation}
L= \theta_{\hat c}^3 - {\hat c} \left(\theta_{\hat c}^3 +
\frac{3}{2}\theta_{\hat c}^2 + \frac{1}{2}\theta_{\hat c} +
2r ( \theta_{\hat c} + \frac{1}{2}) \right),
\label{43}
\end{equation}
with $r=\frac{N-1}{2N^2}$ and $N=6,4,3,2$ respectively.
Since by construction ${\cal C}^{CY}$ can not distinguish between
the mentioned four spaces, it is natural to expect that
(\ref{41}) is related to the common part of their Picard-Fuchs
equation. Indeed, it is easy to see that (\ref{41}) is solution of
\begin{equation}
L_1= \theta_{\hat c}^3 - {\hat c} \left(\theta_{\hat c}^3 +
\frac{3}{2}\theta_{\hat c}^2 + \frac{1}{2}\theta_{\hat c} \right)
- \frac{1}{4}\theta_{\hat c}.
\label{44}
\end{equation}
The operators $L$ and $L_1$ differ in the first order differential
operator
\begin{equation}
L_2= L-L_1= \frac{1}{4}\theta_{\hat c} - 2r {\hat c}
\left(\theta_{\hat c} + \frac{1}{2} \right).
\end{equation}

The singular points of the $K3$ Picard-Fuchs equation (\ref{43})
are at $\hat{c}=0,1,\infty$. At $\hat{c}=1$ the
$K3$ develops an $A_1$ singularity, and its associated monodromy
reproduces the Weyl group for $SU(2)$.
However the point $\hat{c}=0$ presents logarithmic monodromy, as
can be deduced from the corresponding indicial equations.
This fact does not allow to use the solutions of (\ref{43}), in
particular the period carrying Weyl monodromy around $\hat{c}=1$,
for defining the Seiberg-Witten differential of a Riemann surface.
Definition (\ref{44}) is the necessary modification of $L$ in order
to achieve this. The singular points of $L_1$ are still at
$\hat{c}=0,1,\infty$, but the asymptotic behavior for both $0,\infty$
have been modified. The indicial equations of $L$ and $L_1$ at
$\hat{c}=1$ coincide, therefore the leading behavior of the solutions
at this singularity is not affected.

We want now to analyze how working with $L_1$ instead of $L$ affects
the physics that we obtain. Namely, we will compare the physics
associated to ${\cal C}^{CY}$ with that of a type IIA string
compactified
in manifolds $A$, $B$, $C$ or $D$, and their corresponding heterotic
dual strings.

Let us begin considering the conifold locus of these Calabi-Yau
manifolds. This locus corresponds to the melting of the
$K3$-fiber singular points $e_{1}^{\pm}$, given in (\ref{3}).
At these points $\hat{c}=1$ and $L$ and $L_1$ coincide.
Therefore, in a neighborhood of the conifold locus,
${\cal C}^{CY}$ (equivalently an $N=4$ softly broken gauge theory)
should describe essentially the same physics as the threefold.
Indeed in both cases the singularity is interpreted as due to BPS
dyons becoming massless \cite{SW,SW2}, \cite{S}.

The singular locus $\hat{{\cal C}}_0$ of the $N=4$ softly broken
theory corresponds to an electric hypermultiplet acquiring
zero mass \cite{SW2}. This multiplet derives from components of the
initial massive adjoint hypermultiplet, and the singularity occurs
for Higgs expectation values of order $m$, with $m$ the adjoint
hypermultiplet bare mass. The gravity counterpart
of this locus is ${\cal C}_0$ in (\ref{eq:c5}), which we observed can
be
represented by the melting of the singular points $e_{0}^{\pm}=
e_{1}^{\pm}$. Since over $e_{0}^{\pm}$ the $K3$ develops the
$\hat{c}=0$
singularity, at which the operators $L$ and $L_1$ strongly differ,
the interpretation of field theory and string loci, $\hat{{\cal
C}}_0$
and ${\cal C}_0$, can naturally be different.
However we will propose that both share a common origin in the
presence
of an additional (non-moduli) scale in the theory. For the $N=4$
softly
broken theory, this scale is of course provided by the mass of the
adjoint
hypermultiplet. In the case of the Calabi-Yau space we want to argue
that
the scale behind ${\cal C}_0$ is the string tension $\alpha'$.
Let us stress that the identification between $m^2$ and
$(\alpha')^{-1}$
was one of the main consequences of the map (\ref{14})
between gauge theory variables $(\hat{u},\tau)$ and string moduli
$(b,c)$.

We will concentrate in the Calabi-Yau model $A:~
\IP_{\{1,1,2,2,6\}}^{4}[12]$. The locus ${\cal C}_0$ was defined as
the set of points invariant under ${\cal A}^2$, with ${\cal A}$ given
by
the moduli space symmetry (\ref{eq:c11}). This transformation
satisfies
${\cal A}^6=-1$, discarding its origin in an additional massless
particle.
The mirror map relates complex structures of $A^{\ast}$,
parameterized
by $(b,c)$, to K\"ahler structures of $A$, parameterized by special
coordinates $(t_1,t_2)$ ($t_1 \sim \mbox{log}~c$, $t_2 \sim
\mbox{log}~b$,
for $c,b \rightarrow 0$ \cite{MM}). Heterotic-type II duality further
relates
the K\"ahler coordinates $(t_1,t_2)$ to the heterotic moduli
\cite{KV,KKLMV}
\begin{equation}
t_1=T~~, \hspace{2cm} t_2=-\frac{S}{2 \pi i},
\label{800}
\end{equation}
with $T$ an space-time modulus and $S$ the heterotic dilaton. In
\cite{can,Y2,KKLMV} the monodromy group for the manifold $A$ was
worked out, from where it can be deduced
\begin{equation}
{\cal A}^2 : T \rightarrow -\frac{T+1}{T}~, \hspace{1cm}
S \rightarrow S.
\label{501}
\end{equation}
We see that the monodromy around ${\cal C}_0$ produces a $T$-duality
transformation for the heterotic dual string\footnote{The monodromy
around the locus ${\cal D}_{(0,-1)}= \{ c=0 \}$ corresponds to $T
\rightarrow
T+1$. Combining this and (\ref{501}), or equivalently surrounding the
two conifold branches, we can obtain the standard $T \rightarrow
-1/T$.}.

More in general, and without doing reference to string-string
dualities,
we can relate ${\cal C}_0$-type loci, i.e. set of fixed points under
global symmetries in moduli space, to self-dual regions with
respect to the Calabi-Yau generalization of the target-space duality
$R\rightarrow \frac{1}{R}$, where $R$ stands
for a K\"ahler modulus. This can can already be observed in the
simpler
case of the quintic \cite{COGP}. In that case we have a single
moduli, $\psi$; the conifold singularity is located at $\psi=1$.
The symmetry transformation ${\cal A}:\psi \rightarrow \beta
\psi$, with $\beta^5=1$, lets fixed the point $\psi=0$, which is the
analog
in this simple example of the ${\cal C}_0$ locus. At this point the
Zamolodchikov's metric is regular, while the K\"{a}hler potential
becomes
singular. However, this singularity in the K\"{a}hler potential is
not
producing any singular contribution to the partition function, as can
be
seen from the topological analysis of reference \cite{BCOV}. Thus, we
should
not expect new massless particles at $\psi=0$\footnote{It must
be recalled that the K\"{a}hler potential enters always into the
special
geometry equations in the form $e^{-K} \| f \|^2$, with
$f$ a meromorphic section of the special geometry Hodge line bundle.
A zero
of $f$ at the origin ($\psi=0$ for the quintic) regularizes the
singularity
of $K$. The meromorphic section $f$ contributes to the definition
of the topological propagator from which the topological version
of the partition function is built up. At the conifold locus, this
is the propagator reproducing the logarithmic singularity at one
loop. At the
${\cal C}_0$ locus, this propagator is smoothed out with the help
of special geometry structure, namely, the freedom to normalize the
vacuum section of the Hodge line bundle.}. Alternatively, the ${\cal
A}$
transformation acts on the Calabi-Yau radius in a similar (but more
complicated) way to $R \rightarrow \frac {1}{R}$ \cite{COGP}.

$T$-duality transformations always imply the introduction of an
additional
scale. Having identified ${\cal C}_0$ as the Calabi-Yau
generalization of
a self-dual locus, we should determine the scale that is associated
to it.
Expression (\ref{501}), together with the fact that the Calabi-Yau
weak
coupling limit $b \rightarrow 0$ gets mapped by string-string duality
to
heterotic perturbative effects \cite{KV}, indicate that this scale
is given by the heterotic-dual string tension $\alpha'$. Restoring
unities,
the self-dual point of (\ref{501}) is $T=\rho~ \sqrt{\alpha'}$, with
$\rho^3
=1$. Therefore perturbative string excitations will
have, at that point, a typical mass ${\alpha'}^{-1/2}$. This agrees
with
the field theory interpretation ${\alpha'}^{-1/2}\sim m$, for $m$ the
adjoint
hypermultiplet bare mass.
Values $b>0$ will imply corrections of the characteristic mass of
string excitations at ${\cal C}_0$. In this case, the natural
candidate
for comparing square masses should be the Casimir expectation value
$u\sim m^2 e_1(\tau)$, that determines the field theory locus
$\hat{{\cal C}}_0$.

{}From the type II perspective, we should obtain the same result for
the typical mass of excitations in a neighborhood of ${\cal C}_0$, by
looking at BPS states associated to Ramond-Ramond charged branes.
Notice that, although ${\cal C}_0$ is equally a self-dual locus on
the
type IIA side, the mass of these excitations will not be governed by
the
type II string tension, since Ramond-Ramond charged branes are
non-perturbative objects \cite{P}.

The same considerations apply to models $B$ and $C$. The case $D$ is
however different, as the monodromy around ${\cal C}_0$ involves
logarithms\footnote{The indicial equations derived from the
third order Picard-Fuchs operator (\ref{C1}) of model $D$, around
$c=\infty$, posses a triple solution.} and therefore it does not
admit to be interpreted as a self-dual locus.
In this case we could think in the appearance of a massless particle
to explain the singularity. If underlying scale
is still $\alpha'$, it would indicate an electrically charged
particle. Physically, the $D$ model can be the one more
closely related to ${\cal C}^{CY}$. It is worth noticing that the
singular
points of the first order differential operator $L_2$ are at
$\hat{c}= 0, 8r,\infty$. Only for the model $D$ we have $8r=1$, and
the
singular points of $L_2$ coincide with those of $L$ and $L_1$.
This fact could be directly connected with passing from a
logarithmic monodromy transformation around the hypermultiplet locus
$\hat{{\cal C}}_0$ in the field theory approach, to an smoothed out
(non-logarithmic) monodromy ${\cal A}^2$ for the models $A$, $B$ and
$C$,
while not for the Calabi-Yau $D$.

Finally we consider the locus ${\cal D}_{(0,-1)}= \{ c=0 \}$, which
correspond to the degenerate situation $e_{1}^+=0$ and
$e_{1}^-=\infty$.
The operators $L$ and $L_1$ also differ at $z=0,\infty$, implying
that
close to this locus ${\cal C}^{CY}$ will not reproduce the string
dynamics. This is indeed as expected, since the mirror map \cite{can}
fixes $t_1=i \infty$ at ${\cal D}_{(0,-1)}$, therefore representing a
decompactification limit. Using (\ref{14}), this locus corresponds to
the
ultraviolet regime $u=\infty$ for the $N=4$ softly broken
theory, where string and field theory should strongly differ.

Summarizing, we have seen that ${\cal C}_{CY}$ can provide a good
description of string phenomena only in a neighborhood of the
conifold
locus. However, ${\cal C}_{CY}$ proved useful in interpreting the
locus
${\cal C}_0$. In the next section, following this path, we will use
${\cal C}_{CY}$ as a tool for understanding further differences
in the coupling to gravity of $SU(2)$ $N=2$ Yang-Mills provided by
models
$A$, $B$, $C$ and $D$.

\section{Higher Kac-Moody Level String Models.}

The differences between the four Calabi-Yau manifolds we are
considering
can be resumed in the properties of the mirror map for their
$K3$-fiber,
or equivalently, the mirror map between $c$ and $t_1$ for the dilaton
modulus value $b=0$. Using the string-string identification $t_1=T$,
we
have
\begin{equation}
c=\frac{h_k(T_{0}^{(k)})}{h_k(T)},
\label{801}
\end{equation}
with $h_k$ the Hauptmodul function of $\Gamma_0 (k)_+$ (shifted by a
constant), and $k=1,2,3,4$ for spaces $A$, $B$, $C$ and
$D$ respectively \cite{can,LY}. The value $T_{0}^{(k)}$ is given by
the
self-dual point of one distinguished element of the group $\Gamma_0
(k)_+$,
the Atkin-Lehner involution
\begin{equation}
T \rightarrow -\frac{1}{kT}.
\end{equation}
{}From (\ref{801}) we see that the $\Gamma_0 (k)_+$ determines the
$T$-duality group of the heterotic dual string. The dual for the $A$
model is a rank three heterotic string compactified in $K3 \times
T^2$
\cite{KV}, where $T$ corresponds to a $T^2$ modulus. Since
$\Gamma_0 (1)_+=PSL(2,Z)$ and $h_1$ coincides with the 
$j$-invariant, we obtain the usual $T$-duality group
for this model.

In \cite{AP} the case $B:\IP^{4}_{\{1,1,2,2,2\} }[8]$ was considered.
It
was proposed to be dual to an heterotic string compactification with
$SU(2)$ perturbative enhancement of symmetry realized at Kac-Moody
level 2. The mirror map (\ref{801}) fixes the heterotic enhancement
of
symmetry point at $T=i/\sqrt{2}$.
The right and left moving momenta for the rank three models are
given \cite{N} by
\begin{eqnarray}
\label{50}
p_L & = & \frac{i \sqrt{2}}{T-\bar{T}}(n_1 + n_2 {\bar T}^2 + 2m
{\bar T}), \nonumber \\
p_R & = & \frac{i \sqrt{2}}{T-\bar{T}}(n_1 + n_2 T \bar{T} +
m (T+\bar{T}) ).
\end{eqnarray}
The condition for new massless states, $p_L=0$, $|p_{R}|^2 \leq 2$,
is satisfied at $T=i/\sqrt{2}$ for $n_1=2 \: n_2=\pm 1/2$, $m=0$.
This implies the value $p_{R}^2 =1$. If we denote the Kac-Moody level
at which the enhanced gauge group is realized by $k_G$, the
following relation holds
\begin{equation}
\label{900}
k_G~ |p_{R}|^2 = 2 .
\end{equation}
Therefore $k_G=2$ for the heterotic dual associated to the manifold
$B$. In an analogous way we can analyze the heterotic duals that
could
correspond to models $C$, $D$. The mirror map fixes the perturbative
enhancement of symmetry at $T=i/\sqrt{k}$, with $k=3,4$ respectively.
At this value, using (\ref{50}), the conditions for two additional
massless states are again verified, satisfying (\ref{900}) for
$k_G=k$.
Although further checks along the lines of \cite{KT,K3,AP} have to be
done for the cases $C$, $D$, and the explicit construction of the
heterotic dual \cite{KV} accomplished for models $B$, $C$ and $D$,
we will assume that a main difference between the coupling to gravity
of $N=2$ $SU(2)$ Yang-Mills that the four manifolds can provide is
given by the Kac-Moody level at which they realize the gauge
symmetry,
being $k_G=1,2,3,4$ respectively.

The Kac-Moody level also affects the relation between gauge group
bare
coupling constant and heterotic dilaton \cite{G}. With a convenient
normalization we can set $\tau_G=- \frac{k S^{inv}}{2 \pi i}$, where
now it is important to distinguish between the special coordinate
dilaton $S$, with modular properties under $T$-duality
transformations,
and the invariant dilaton $S^{inv}$ \cite{Sinv,Sinv2}.
Since ${\cal C}^{CY}$ captures the Calabi-Yau information in a
neighborhood of the conifold locus, and in the limit $b\rightarrow 0$
this information reproduces the heterotic enhancement of symmetry, we
can, in this limit, identify $\tau$ and $\tau_G$. Using (\ref{15}),
the
relation between $\tau$ and $S^{inv}$ translates into
\begin{equation}
t_2 = - \frac{kS}{2 \pi i}
\label{51}
\end{equation}
for the special geometry coordinates.

We have analyzed properties of the models $A$, $B$, $C$
and $D$ which correspond to the heterotic-dual weak coupling regime.
We
can now use both pieces of information, namely ${\cal C}^{CY}$ and
the
level $k$, in trying to reproduce properties associated to the
heterotic
strong coupling limit. We will concentrate therefore on the locus
${\cal C}_1=\{ b=1 \}$, which by the mirror map implies $t_2=0$
\cite{KM}.
On the type IIA side this locus is associated to the appearance of a
curve of {\bf Z}$_2$ quotient singularities, ${\cal C}_{Z_2}$,
corresponding to blowing down the exceptional divisor whose size
controls
the dilaton modulus $b$. In \cite{KMP,KM} this singularity was
interpreted
as a non-perturbative enhancement of symmetry in the $U(1)$ vector
field
associated to the dilaton, $U(1) \rightarrow SU(2)$, together with
the
appearance of massless hypermultiplets in the adjoint representation
of
$SU(2)$. The number of massless adjoint hypermultiplets is given by
the
genus $g$ of the curve of singularities ${\cal C}_{Z_2}$. The
monodromy around ${\cal C}_1$ can be resumed in the $2\times 2$
matrix
\begin{equation}
\left(
\begin{array}{cc}
-1 & 2(g-1) \\
 0 &  -1
\end{array}
\right),
\label{52}
\end{equation}
acting on a column vector whose first entry we denote by $t_{2}^D$,
and
the second is given by the special coordinate $t_2$.

Since $t_2$ behaves as the special coordinate version of the $N=4$
coupling constant $\tau$, and $\tau$ is an intrinsic parameter of
${\cal C}^{CY}$, it should be possible to describe the different
monodromy matrices (\ref{52}) for the spaces $A$, $B$, $C$ and $D$
in an unified way. In order to do this,
we will allow the freedom to change the normalization of $t_{2}^D$,
while preserving that of $t_2$. Inspired by (\ref{51}) we consider
the
change $t_{2}^D \rightarrow t_{2}^D/k$, transforming (\ref{52}) into
\begin{equation}
\left(
\begin{array}{cc}
-1 & 2(g-1)/k \\
 0 &  -1
\end{array}
\right).
\label{eq:pf8}
\end{equation}
For the four models we are considering the following relation between
the genus $g$ of the curve of singularities at ${\cal C}_1$, and the
level $k$ dictated by the modular properties of the mirror map, is
verified (see Appendix C)
\begin{equation}
\frac{g-1}{k}= 1.
\end{equation}
Substituting this into (\ref{eq:pf8}) we obtain a single
representation
of the ${\cal C}_1$ monodromy for the four Calabi-Yau spaces. It is
given
by $P T^{-2} $, where now $T$ denote the $Sl(2;Z)$
generator and $P=-1$.

It must be noticed that a $T^{-2}$ monodromy is the one we will
get for the Donagi-Witten curve at the singular locus
$\tau=0$\footnote{The dependence on $\tau$ of the Donagi-Witten
construction is built using the Weierstrass invariants $e_{i}(\tau)$
(see footnote 3), which are modular functions of $\Gamma(2)$.
Therefore
the monodromy matrices for the singular loci $\tau=0,1,i
\infty$ in (\ref{12}), have to be conjugate to $T^2$ or $T^{-2}$.},
which by the map between field theory and string variables (\ref{14})
corresponds to ${\cal C}_1$. Namely, encoding the monodromy around
$\tau=0$ through its action on a column vector with
entries $1$ and $\tau$, the matrix $T^{-2}$ implies $\tau^D
\rightarrow
\tau^D + 2$, with $\tau^D =-1/\tau$.
We observe that, although the ${\cal C}^{CY}$ curve can not
describe the physics at ${\cal C}_1$ because is missing the parameter
$g$ and the Weyl generator $P$, it can however explain its underlying
structure. Indeed, using ${\cal C}^{CY}$ and the Kac-Moody level
information, it is possible to reproduce in a natural way (\ref{52}).
We can also put in correspondence the monodromies for the
Calabi-Yau and field theory loci ${\cal C}_{\infty}$ and $\tau=
i\infty$.
Taking into account the double covering of the string moduli space
that the field theory is doing, the monodromy $S T^{-2} S^{-1}$ 
around $\tau =
i \infty$, producing $\tau \rightarrow \tau +2$, becomes the
${\cal C}_{\infty}$ monodromy \cite{can,Y2} $t_2 \rightarrow t_2 +
1$.

\section{Final Comments.}

Ultraviolet finite $N=2$ supersymmetric gauge theories are described
\cite{SW2} in
terms of the Higgs vacuum expectation values, the bare coupling
constant
$\tau$ and the masses of their matter content. In order to relate
these
theories with a string compactification on a Calabi-Yau threefold,
we can not follow the standard path of performing the heterotic point
particle limit. In this paper we have shown that $N=2$ gauge theories
with $N=4$ matter content
are associated with the global structure of $K3$-fibration
threefolds. Moreover the global geometrical
information on $K3$-fibrations reproduces all the ingredients used in
the characterization
\cite{DW} of the integrability of these ultraviolet finite $N=2$
theories.
For the case of $SU(2)$,
our philosophy has been to consider the gauge theory with $N=4$
matter
as encoding the common geometrical structure of $K3$-fibrations
sharing the same $h_{2,1}$ Hodge number, and structure of the complex
deformations moduli space.

The special role of a ultraviolet finite $N=2$ $SU(2)$ gauge theory,
in particular the case $2N_c = N_f$, appears in a different context
\cite{Sen} as proves \cite{BDS} of F-theory \cite{V} backgrounds. The
prove approach is related to duality properties of type II strings,
while our approach should be related to strong-weak coupling duality
of the heterotic string.

\begin{center}
{\bf Acknowledgments}
\end{center}

\vspace{2 mm}

We thank conversations with K. Landsteiner, W. Lerche, A. Sen, C.
Vafa and
N. Warner. This work is partially supported under grant OFES
contract number 930083 by European Community grant ERBCHRXCT920069
and by
PB92-1092. The work of E. L. is supported by C.A.P.V. fellowship. The
work
of R. H. is supported by U.A.M. fellowship.

\vspace{1.5cm}

\noindent
{\Large \bf Appendices.}

\appendix

\section{Isogeny Between Cubic and Quartic Curves.}

We consider an elliptic curve, given by the quartic
\begin{equation}
y^2= (x^2 + a)^2 - b^2,
\label{a1}
\end{equation}
for certain constants $a$ and $b$. By defining
$x'=x^2 + a$, $y'=yx$ we can  convert (\ref{a1}) into an
isogenic cubic
\begin{equation}
{y'}^2=(x'-a)({x'}^2 - b^2).
\label{a2}
\end{equation}
This process amounts to quotienting by the symmetry
$x \rightarrow -x$ of (\ref{a1}), and blowing down the
line $(x=0,y)$ in order to get again an elliptic
curve.

It is immediate to see that the abelian differential for
quartic and cubic are related by
\begin{equation}
\frac{dx'}{y'}=2\frac{dx}{y}.
\end{equation}
Let us choose a basis of cycles of (\ref{a2}) in the following form:
$\gamma_1'$ surrounding the branch points $\pm b$, and $\gamma_2'$
surrounding the branch points $b$ and $-a$. With this
choice the cycles $\gamma_1$ and $\gamma_2$, transformed of
$\gamma_1'$ and $\gamma_2'$ respectively, satisfy
\begin{equation}
\gamma_1=\gamma_1'~~, \hspace{2cm} \gamma_2=2\gamma_2'.
\end{equation}
Therefore the difference between quartic and cubic, from
the point of view of the periods, reduces to a change
in normalization
\begin{equation}
\int_{\gamma_1} \omega = \frac{1}{2} \int_{\gamma_1'} \omega'~~,
\hspace{1cm} \int_{\gamma_2} \omega= \int_{\gamma_2'} \omega',
\label{A1}
\end{equation}
for $\omega$, $\omega'$ the respective abelian differentials.
This is precisely the difference between the two curves considered
in \cite{SW}, \cite{SW2} for solving $SU(2)$ $N=2$ Yang-Mills.

The quartic form (\ref{a1}) reproduces the structure, for
the particular case $SU(2)$, of the hyperelliptic curves for higher
rank gauge groups proposed in \cite{KLTY,AF}
\begin{equation}
y^2=P_{SU(N)}(x;u_i)^2 - \Lambda^{2N},
\end{equation}
with $u_i$ the $SU(N)$ Casimirs.

We should also note that the Jacobi invariants for (\ref{a1})
and (\ref{a2}) differ. This can be seen by transforming both
curves to the standard form
\begin{equation}
y^2=x(x-1)(x-\lambda),
\end{equation}
and comparing the $\lambda$-parameters obtained. We have
\begin{equation}
\lambda=\frac{a+\sqrt{a^2 - b^2}}{a -\sqrt{a^2 - b^2}}~~,
\hspace{1cm} \lambda'=\frac{a+b}{a-b}.
\end{equation}

\section{The Map Between $SU(3)$ Moduli Spaces.}

In \cite{KV} the Calabi-Yau $\IP_{\{1,1,2,8,12\}}^{4}[24]$ was
proposed as the type II dual for the string embedding of $N=2$
$SU(3)$ Yang-Mills theory. In this Appendix, extending part of the
analysis done in previous sections for $SU(2)$, we will build a map
between the moduli
space of $N\!=\!2$ $SU(3)$ Yang-Mills theory with adjoint matter and
the moduli space of the hypersurface $\IP_{\{1,1,2,8,12\}}^{4}[24]$.
The analysis will be restricted to the weak coupling sections
$\tau =i \infty$ and $b=0$ of the respective moduli spaces.

The discriminant locus of the genus two curve describing $SU(3)$ with
adjoint matter \cite{DW}, in the limit $\tau=i \infty$, is given by
\begin{equation}
\Delta_{SU(3)} = (4\hat{u}^3-27\hat{v}^2)(4(\hat{u}-1)(\hat{u}-4)^2-
27\hat{v}^2),
\label{eq:b1}
\end{equation}
where $\hat{u}$ and $\hat{v}$ are the quadratic and cubic Casimir
expectation values respectively, normalized by the adjoint
hypermultiplet
mass. The first factor corresponds to a classical enhancement of
$SU(2)$
gauge symmetry. We will denote its two branches by
\begin{equation}
{\cal C}^{\pm} = \{3\sqrt{3} \hat{v} = \pm 2 \hat{u}^{3/2} \}.
\label{eq:b2}
\end{equation}
The second factor goes to zero when some component of the $N\!=\!4$
hypermultiplet become massless; we will also divide it into
\begin{equation}
{\cal C}_{h}^{\pm} = \{3\sqrt{3} \hat{v} = \pm
2(\hat{u}-4)\sqrt{\hat{u}-1} \}.
\label{eq:b3}
\end{equation}

If we now concentrate on the $K3$-fibration structure of the
$\IP_{\{1,1,2,8,12\}}^{4}[24]$ Calabi-Yau manifold \cite{KLMVW}
\begin{eqnarray}
W & = & \frac{1}{24} (z + \frac {b}{z} + 2) +
\frac{1}{12} x_{3}^{12} + \frac{1}{3} x_{4}^3 + \frac{1}{2} x_{5}^2 +
\nonumber \\
&&  + \frac {1}{6\sqrt{c}}(x_0 x_3)^6 + \left( \frac {a}{\sqrt{c}}
\right)^{1/6} x_0 x_3 x_4 x_5 = 0,
\end{eqnarray}
we notice that the $K3$-fiber becomes singular at six points
$(z=e_i^{\pm},
\: \: i=0,1,2)$ of the $\IP^1$ base space
\begin{eqnarray}
e_0^{\pm} & = & -1 \pm \sqrt{1-b}, \nonumber\\
e_1^{\pm} & = & \frac {1-c \pm \sqrt{(1-c)^2-bc^2}}{c}, \nonumber\\
e_2^{\pm} & = & \frac {(1-a)^2-c \pm \sqrt{((1-a)^2-c)^2-bc^2}}{c}.
\label{eq:b5}
\end{eqnarray}
Merging of these points leads to two conifold-type loci, when
$e_i^{+}=
e_i^{-}$ for $i=1,2$, and a strong dilaton locus ${\cal C}_1$, when
$e_0^{+}=e_0^{-}$:
\begin{eqnarray}
{\cal C}_C^{(1)} & = & \{ (1-c)^2-bc^2 = 0 \}, \nonumber\\
{\cal C}_C^{(2)} & = & \{ ((1-a)^2-c)^2-bc^2 = 0 \}, \nonumber\\
{\cal C}_1       & = & \{ b-1 = 0 \}.
\label{eq:b6}
\end{eqnarray}
The weak gravity locus
\begin{equation}
{\cal C}_{\infty} = \{ b = 0 \},
\end{equation}
corresponds to $0=e_0^+=e_1^+=e_2^+$. We can also consider the
merging
$e_0^{\pm}=e_i^{\pm}, \: \: i=1,2$, which happens respectively at
\begin{eqnarray}
{\cal C}_0^{+} & = & \{ c = \infty ~~~ (\forall~ a/\sqrt{c} ) \},
\nonumber \\
{\cal C}_0^{-} & = & \{ a -1 = 0 \},
\label{eq:b7}
\end{eqnarray}

We will restrict now to the $(a,c)$ section of the Calabi-Yau
moduli space at $b=0$. In the spirit of (\ref{14}), we can propose
the
following map between the $(u,v)$-plane at $\tau= i \infty$, and the
Calabi-Yau moduli space at $b=0$
\begin{equation}
1-c = \frac {-2\hat{u}^{3/2}+3\sqrt{3}\hat{v}}{2(\hat{u}-4)
\sqrt{\hat{u}-1}+3\sqrt{3}\hat{v}}, \: \: \: \:
1- \frac {c}{(1-a)^2} = \frac {2\hat{u}^{3/2}+3\sqrt{3}
\hat{v}}{2(\hat{u}-4)\sqrt{\hat{u}-1}-3\sqrt{3}\hat{v}}.
\label{eq:b8}
\end{equation}
This map is built by requiring that the loci ${\cal C}^{\pm}$
get mapped into the two conifold ${\cal C}_{C}^{(1,2)}$, and the
hypermultiplet loci ${\cal C}_{h}^{\pm}$ go into ${\cal C}_0^{\pm}$.

The point $(a=0,c=1)$, one of the intersections between
${\cal C}_C^{(1)}$ and
${\cal C}_C^{(2)}$ at $b=0$, corresponds by heterotic-type II
duality to a perturbative $SU(3)$ enhancement of symmetry point
\cite{KV}. A first check of (\ref{eq:b8}) is that, sending
the mass of the adjoint hypermultiplet $m \rightarrow \infty$ and
identifying again $m^2$ with the string tension $(\alpha')^{-1}$,
it reproduces the point particle limit map for $SU(3)$
\cite{KKLMV,KLMVW}
\begin{eqnarray}
a & = & -2 (\alpha u)^{3/2}, \nonumber\\
c & = & 1 - \alpha^{3/2}(-2 u^{3/2} +3\sqrt{3} v).
\end{eqnarray}
The scale governing the strong coupling effects of the $N=2$ $SU(3)$
Yang-Mills theory with adjoint matter is given by
$\Lambda^{6} \sim e^{2 \pi i \tau} m^{6}$ \cite{DW}. Assuming the
relation (\ref{15}) between the dilaton modulus $b$ and the $N=4$
coupling constant $\tau$, this implies the double scaling
identification
required to perform the point particle limit \cite{KKLMV}
\begin{equation}
b = \alpha^{3} \Lambda^{6}.
\end{equation}

It would be interesting to see if the map (\ref{eq:b8}) can help to
locate, by comparison with the field theory, Argyres-Douglas points
\cite{AD} in string theory.

\section{Summary of $A$, $B$, $C$ and $D$ models.}

The Picard-Fuchs equations that govern the K\"ahler structure
deformations
of the Calabi-Yau manifolds $A:\IP_{\{1,1,2,2,6\}}^{4}[12]$,
$B:\IP_{\{1,1,2,2,2\}}^{4}[8]$, $C:\IP_{\{1,1,2,2,2,2\}}^{5}[4,6]$
and
$D: \IP_{\{1,1,2,2,2,2,2\}}^{6}[4,4,4]$, consist of two differential
operators of second and third order \cite{CYY,Y2}. The second order
operator is common to the four manifolds
\begin{equation}
L^{(1)}= 4\theta_b^2 - b (2\theta_b - \theta_c + 1)
(2\theta_b - \theta_c),
\label{c1}
\end{equation}
while they differ in the third order one
\begin{eqnarray}
\label{C1}
A: && L^{(2)}= \theta^2_c \: (\theta_c - 2\theta_b) - c(\theta_c 
+ 5/6)(\theta_c + 1/2)(\theta_c + 1/6), \nonumber \\
B: && L^{(2)}= \theta^2_c \: (\theta_c - 2\theta_b) - c(\theta_c 
+ 3/4)(\theta_c + 1/2)(\theta_c + 1/4), \nonumber \\
C: && L^{(2)}= \theta^2_c \: (\theta_c - 2\theta_b) - c(\theta_c 
+ 2/3)(\theta_c + 1/2)(\theta_c + 1/3),  \\
D: && L^{(2)}= \theta^2_c \: (\theta_c - 2\theta_b) - c(\theta_c 
+ 1/2)^3. \nonumber
\end{eqnarray}
We are using the normalization conventions for the moduli
parameters $(b,c)$ implied in (\ref{1000}) and (\ref{2}).
The second order operator (\ref{c1}), in the point particle limit
defined by (\ref{eq:cc1}), becomes precisely 
the Picard-Fuchs operator
for $N=2$ $SU(2)$ Yang-Mills theory \cite{KKLMV}.

These four spaces develop a curve of {\bf Z}$_2$ quotient
singularities,
${\cal C}_{Z_2}$, at ${\cal C}_1=\{ b=1 \}$. This singularity comes
from an ambient singularity in the weighted projective spaces in
which
they are immersed. Namely, in our case points
$(x_1,x_2,x_3,x_4,..)$ and $(\lambda x_1, \lambda x_2, \lambda^2 x_3,
\lambda^2 x_4,..)$ should be identified, fixing, for $\lambda=-1$,
the
hyperplane of orbifold points $x_1=x_2=0$. The intersection of this
hyperplane with the defining hypersurface, or complete intersection,
of
the Calabi-Yau manifolds determines ${\cal C}_{Z_2}$ in each case.
The
genus of ${\cal C}_{Z_2}$ is $g=2,3,4,5$ for the spaces $A$, $B$, $C$
and
$D$ respectively \cite{can,KMP}.

The exceptional divisor whose size controls the dilaton modulus $b$,
is introduced in order to resolve the mentioned singularity. However
for the value $b=1$, the manifolds develop the original curve of
{\bf Z}$_2$
quotient singularities. This structure can be read from the following
piece in the defining expression of their mirror manifolds
\begin{equation}
W^{\ast}= \frac{1}{2n} x_1^{2n} + \frac{1}{2n} x_2^{2n} +
\frac{1}{\sqrt{b}~n} (x_1 x_2)^n +...~.
\end{equation}
Changing variables to $x_1/x_2= z^{1/n} b^{-1/2n}$ and $x_1^2= x_0
z^{1/n}$
\cite{KLMVW}, the $K3$-fibration structure (\ref{999}) for
manifolds $A$, $B$, $C$ and $D$ becomes manifest.
The $K3$-fiber of their mirror, which for model $A$ is given in
(\ref{1001}), is given by
\begin{eqnarray}
B: &&   x_{0}^{4}/4 + x_{3}^4/4 +
x_{4}^4/4 + x_{5}^4/4
+ \hat{c}^{-1/4}  x_0 x_3 x_4 x_5 =0, \frac{}{} \nonumber  \\
C: && x_{0}^2/2 + x_{3}^2/2
+ x_{4}^2/2 + \hat{c}^{-1/5}  x_5 x_6 =0 \nonumber \\
&& x_{5}^3/3 + x_{6}^3/3 +\hat{c}^{-1/5} x_0 x_3
x_4 =0 \frac{}{}, \nonumber \\
D: && x_{0}^2/2 + x_{3}^2/2
+ \hat{c}^{-1/6}  x_4 x_5 =0 \nonumber \\
&& x_{4}^2/2 + x_{5}^2/2 + \hat{c}^{-1/6}
x_6 x_7=0 \nonumber \\
&& x_{6}^2/2 + x_{7}^2/2 + \hat{c}^{-1/6}  x_0 x_3=0,
\end{eqnarray}
with $\hat{c}(z;b,c)$ defined in (\ref{557}). The Picard-Fuchs
equation
governing the above $K3$ surfaces can be obtained from the third
order
differential operator of the threefold (\ref{C1}), by sending
$b \rightarrow 0$.

\begin{center}
\begin{tabular}{|clccc|}      \hline
Model    & Manifold           & $h_{2,1}$~~ & Genus $g$ & Mirror  \\
	 &                 &           &       & modular group      \\
$A$      & $\IP_{\{1,1,2,2,6\}}^{4}[12]$    &  $128$    & $2$   &
$\Gamma$ \\
$B$      & $\IP_{\{1,1,2,2,2\}}^{4}[8]$     &  $86$   & $3$   &
$\Gamma_0(2)_+$   \\
$C$      & $\IP_{\{1,1,2,2,2,2\}}^{5}[4,6]$   &  $68$      & $4$   &
$\Gamma_0(3)_+$    \\
$D$      & $\IP_{\{1,1,2,2,2,2,2\}}^{6}[4,4,4]$~~ &  $58$      & $5$
 &
$\Gamma_0(4)_+$   \\  \hline
\end{tabular}
\end{center}

The four manifolds we are considering have common Yukawa couplings
\cite{CYY,Y2}
\begin{equation}
K_{ccc}= \frac{1}{c^3 \Delta_C},~ K_{ccb}=\frac{1-c}{2 c^2 b
\Delta_C},~
K_{cbb}=\frac{2c-1}{4cb\Delta_C \Delta_1},~ K_{bbb}= \frac{1-c+b-
3cb}{8b^2 \Delta_C \Delta_1^2},
\end{equation}
where $\Delta_C=(1-c)^2 - c^2 ~ b$ is the conifold factor in the
discriminant, and $\Delta_1=1-b$. To finish, let us notice that the
Yukawa couplings of Calabi-Yau three-folds are determined by the
Picard-Fuchs equations coefficients $f_{l}^{k_1,..,k_n}$
\begin{equation}
L^{(l)} = \sum_{k_1,..,k_n} f_{l}^{k_1,..,k_n} \partial_{c_1}^{k_1}..
\partial_{c_n}^{k_n}~~,
\end{equation}
with $\sum k_i \geq 2$ \cite{CYY} ($c_i$, $i=1,..,n$ includes all
Calabi-Yau moduli parameters and $l$ labels the set of Picard-Fuchs
differential operators). This information is retained in the
modified Picard-Fuchs operator (\ref{44}) that governs the
Seiberg-Witten differential for ${\cal C}^{CY}$.


\end{document}